\documentclass[acmsmall]{acmart}
\usepackage{CJKutf8}
\usepackage{multirow}
\usepackage{amsmath,amsfonts}
\usepackage{algorithmic}
\usepackage{graphicx}
\usepackage{booktabs}
\usepackage{subfigure}
\usepackage{textcomp}
\usepackage{xcolor}
\usepackage{color}
\usepackage{comment}
\usepackage{bibentry}
\usepackage{url}
\usepackage{makecell}
\usepackage{longtable}
\usepackage{lipsum}
\AtBeginDocument{%
  \providecommand\BibTeX{{%
    \normalfont B\kern-0.5em{\scshape i\kern-0.25em b}\kern-0.8em\TeX}}}

\setcopyright{acmlicensed}
\copyrightyear{2018}
\acmYear{2018}
\acmDOI{XXXXXXX.XXXXXXX}
\acmConference[ACM IH\&MMSec'25]{13th ACM Workshop on Information Hiding and Multimedia Security}{June 18--20, 2025}{San Jose, CA, USA}

\begin{document}

\title{AgriSentinel: Privacy-Enhanced Embedded-LLM Crop Disease Alerting System}

\author{Honghui Xu}
\affiliation{%
	\institution{Department of Information Technology, Kennesaw State University}
	\city{Marietta}
	\state{Georgia}
	\country{USA}
	\postcode{30060}
}
\email{hxu10@kennesaw.edu}

\author{Chanti Raju Mylay}
\affiliation{%
	\institution{Department of Computer Science, Kennesaw State University}
	\city{Marietta}
	\state{Georgia}
	\country{USA}
	\postcode{30060}
}
\email{cmylay@students.kennesaw.edu}

\author{Taeyeong Choi}
\affiliation{%
	\institution{Department of Information Technology, Kennesaw State University}
	\city{Marietta}
	\state{Georgia}
	\country{USA}
	\postcode{30060}
}
\email{tchoi3@kennesaw.edu}

\author{Bobin Deng}
\affiliation{%
	\institution{Department of Computer Science, Kennesaw State University}
	\city{Marietta}
	\state{Georgia}
	\country{USA}
	\postcode{30060}
}
\email{bdeng2@kennesaw.edu}

\author{Zhipeng Cai}
\affiliation{%
	\institution{Department of Computer Science, Georgia State University}
	\city{Atlanta}
	\state{Georgia}
	\country{USA}
	\postcode{30303}
}
\email{zcai@gsu.edu}

\renewcommand{\shortauthors}{Xu, et al.}

\begin{abstract}

Crop diseases pose significant threats to global food security, agricultural productivity, and sustainable farming practices, directly affecting farmers’ livelihoods and economic stability. To address the growing need for effective crop disease management, AI-based disease alerting systems have emerged as promising tools by providing early detection and actionable insights for timely intervention. However, existing systems often overlook critical aspects such as data privacy, market pricing power, and farmer-friendly usability, leaving farmers vulnerable to privacy breaches and economic exploitation. To bridge these gaps, we propose AgriSentinel, the first Privacy-Enhanced Embedded-LLM Crop Disease Alerting System. AgriSentinel incorporates a differential privacy mechanism to protect sensitive crop image data while maintaining classification accuracy. Its lightweight deep learning-based crop disease classification model is optimized for mobile devices, ensuring accessibility and usability for farmers. Additionally, the system includes a fine-tuned, on-device large language model (LLM) that leverages a curated knowledge pool to provide farmers with specific, actionable suggestions for managing crop diseases, going beyond simple alerting. Comprehensive experiments validate the effectiveness of AgriSentinel, demonstrating its ability to safeguard data privacy, maintain high classification performance, and deliver practical, actionable disease management strategies. AgriSentinel offers a robust, farmer-friendly solution for automating crop disease alerting and management, ultimately contributing to improved agricultural decision-making and enhanced crop productivity.

\end{abstract}



\maketitle

\section{Introduction}\label{sec:introduction}

Crop diseases are emerging as some of the greatest threats to global food security and agricultural productivity~\cite{ristaino2021persistent}, directly impacting farmers’ livelihoods~\cite{mehta2023revolutionizing}, economic stability~\cite{sharma2024revolutionizing}, and the broader goal of sustainable agriculture. To achieve crop disease alerting, there are increasing approaches to provide reliable tools to enhance farmers’ ability to manage crop health effectively.

AI-based disease alerting systems offer a promising solution by providing early detection of crop diseases and actionable insights for timely intervention. By leveraging advanced AI technologies~\cite{nettleton2019predicting,tantalaki2019data}, such a system empowers farmers to make informed decisions that minimize crop losses, optimize yield, and reduce the reliance on excessive chemical inputs. This proactive approach not only improves productivity but also promotes sustainable agricultural practices by conserving resources, protecting the environment, and supporting long-term farming viability in the face of challenges such as pest outbreaks and changing environmental conditions~\cite{liang2023iot,mishra2021automation}. Moreover, with the emergence of large language models (LLMs)~\cite{xu2024device,chen2024octo,chen2024octopusv2,chen2024octopusv3,chen2024octopusv4}, these models excel at extracting structured information from unstructured data sources, including reports, news articles, and farmer records, providing actionable insights that enhance various agricultural tasks~\cite{banerjee2024study,tzachor2023large}. Furthermore, these models facilitate automated decision-making by integrating with other AI systems, streamlining the identification and management of crops and pests.

Unfortunately, these data-driven AI systems are vulnerable to malicious attacks, such as membership inference attacks~\cite{jalalzai2022membership,chen2020practical}, where attackers exploit the model to determine whether specific data points were used during training. In the context of agriculture, such attacks can expose sensitive crop data, including information about crop health, yield patterns, and regional farming practices~\cite{gupta2020security}. This stolen data can then be used to predict crop production levels, allowing market players to manipulate pricing strategies to their advantage~\cite{sykuta2016big,javaid2022enhancing}. Such actions can disrupt fair market dynamics, leading to profit losses for farmers who are unable to compete effectively or secure equitable prices for their produce. Safeguarding crop data privacy is, therefore, essential to protect farmers’ economic interests and ensure a fair agricultural marketplace.

Although existing studies have explored AI-based crop disease alerting, they often lack a comprehensive focus on critical aspects such as privacy preservation to protect market pricing power and the development of farmer-friendly tools. This highlights the need for more holistic solutions that address both technical challenges and practical usability for farmers. To address this gap, we propose AgriSentinel, a Privacy-Enhanced Embedded-LLM Crop Disease Alerting System designed to achieve three key objectives: First, it incorporates a robust data privacy protection mechanism to safeguard sensitive crop data from potential misuse. Second, it features an on-device LLM-based model that not only provides farmers with specific strategic suggestions for managing crop diseases but also moves beyond simple alerting. Finally, its lightweight deep learning architecture is optimized for mobile devices, offering a more farmer-friendly and accessible solution to improve agricultural decision-making.

To sum up, this paper makes several significant contributions:
\begin{itemize}
    \item To the best of our knowledge, this is the first work to design a privacy-enhanced embedded-LLM crop disease alerting system.

    \item We propose the AgriSentinel framework, which offers farmers a secure, user-friendly, LLM-based application for crop disease alerting and management.

    \item A lightweight, differentially private deep learning-based crop disease classification model is developed within AgriSentinel, enabling accurate disease classification while safeguarding the privacy of crop image data.

    \item In AgriSentinel, an on-device LLM-based model is fine-tuned to provide farmers with specific, actionable suggestions for managing crop diseases, utilizing a curated knowledge pool of crop disease information.

    \item Comprehensive experiments are conducted to validate the effectiveness and practicality of the proposed AgriSentinel framework.
\end{itemize}

The remainder of this paper is organized as follows: Section~\ref{sec:related-work} provides a brief overview of related work, while Section~\ref{sec:methodology} presents a detailed description of the proposed AgriSentinel framework. In Section~\ref{sec:experiment}, we conduct real-world data experiments and analyze the results. Finally, the paper concludes with key insights in Section~\ref{sec:conclusion}.

\section{Related Work}\label{sec:related-work}

This section reviews the related works of crop disease alerting systems, agricultural large language models, and privacy-enhanced AI approaches.

\subsection{Crop Disease Alerting Approaches}\label{subsec:crop_dis_alert}

Crop disease alerting approaches can be broadly categorized into context-aware models and data-driven models.
(1) Context-aware models~\cite{liang2023iot,mishra2021automation} integrate broader contextual information, such as regional agricultural practices and environmental factors. For instance, real-time sensor inputs and historical weather data play a crucial role in improving the accuracy of disease forecasting. Additionally, social media data can provide insights into local agricultural sentiment, offering a holistic understanding of crop health dynamics.
(2) Data-driven models~\cite{nettleton2019predicting,tantalaki2019data} leverage diverse data sources, applying machine learning and deep learning techniques for disease prediction. CNNs have been widely used for classifying crop disease images~\cite{chillakuru2022enhanced}. Techniques like feature-level fusion and ensemble learning enhance predictive accuracy by combining outputs from multiple data-driven models~\cite{jimenez2016observation}. Furthermore, attention mechanisms improve the interpretability of extracted features~\cite{nong2023surveillance}, while representation learning aids in identifying critical patterns for precise crop disease prediction.

\subsection{Agricultural Large Language Models}\label{subsec:agricultural_LLMs}

Modern LLMs like GPT and BERT have revolutionized data processing and generation with their exceptional understanding and adaptability~\cite{sapkota2024multi}. Recently, these models have begun transforming the agricultural sector by addressing complex challenges such as disease detection, crop monitoring, and strategic decision-making~\cite{tzachor2023large}.
In agriculture, LLMs excel at extracting structured information from unstructured data sources, including reports, news articles, and farmer records, providing actionable insights that enhance various agricultural tasks~\cite{banerjee2024study}. Furthermore, these models facilitate automated decision-making by integrating with other AI systems, streamlining the identification and management of crops and pests. Their few-shot learning capabilities are particularly advantageous in agriculture, where data collection is often time-intensive and costly.
However, implementing LLMs in agriculture poses significant challenges. These include the need for high-quality, diverse datasets and the computational resources required to train and deploy the models~\cite{andreev2024text}. Additionally, ethical concerns, such as data privacy and security, add further complexity to their practical adoption~\cite{silva2023gpt}.

\subsection{Privacy-Enhanced AI Approaches}\label{subsec:PPL}

Preserving data privacy is critical in the era of big data, enabling meaningful analysis and model training while safeguarding sensitive information. Four key methodologies are commonly employed for privacy protection in AI. 
(1) Differential privacy, widely adopted in machine learning frameworks, introduces random noise to data or model outputs to prevent the identification of individual data points~\cite{aggarwal2024resource}.
(2) Federated learning enhances privacy by training models across decentralized devices or servers holding local data, eliminating the need for data sharing~\cite{rahman2021comparative}. 
(3) Homomorphic encryption enables computations on encrypted data, ensuring sensitive information remains concealed~\cite{el2024preserving}. 
(4) Secure multi-party computation (SMPC) facilitates privacy-preserving collaborative learning by allowing multiple parties to compute functions over private inputs~\cite{gowda2024secure}. 
As the demand for privacy in data-driven applications grows, the integration of these techniques into AI pipelines is becoming increasingly vital to address risks associated with data breaches~\cite{aouedi2024survey}.

The development of a privacy-enhanced LLM-based crop disease alerting system presents two significant technical challenges. 
First, designing a lightweight and user-friendly system that allows farmers to easily adopt and leverage the transformative capabilities of large language models for effective crop disease management is inherently complex. 
Second, ensuring privacy-enhanced training for agricultural LLMs is critical, as inadequately protected crop data could be exploited to infer production levels, potentially undermining farmers’ market pricing power and causing economic losses.

\section{AgriSentinel Model}\label{sec:methodology}

\begin{figure*}[h]
  \includegraphics[width=\textwidth]{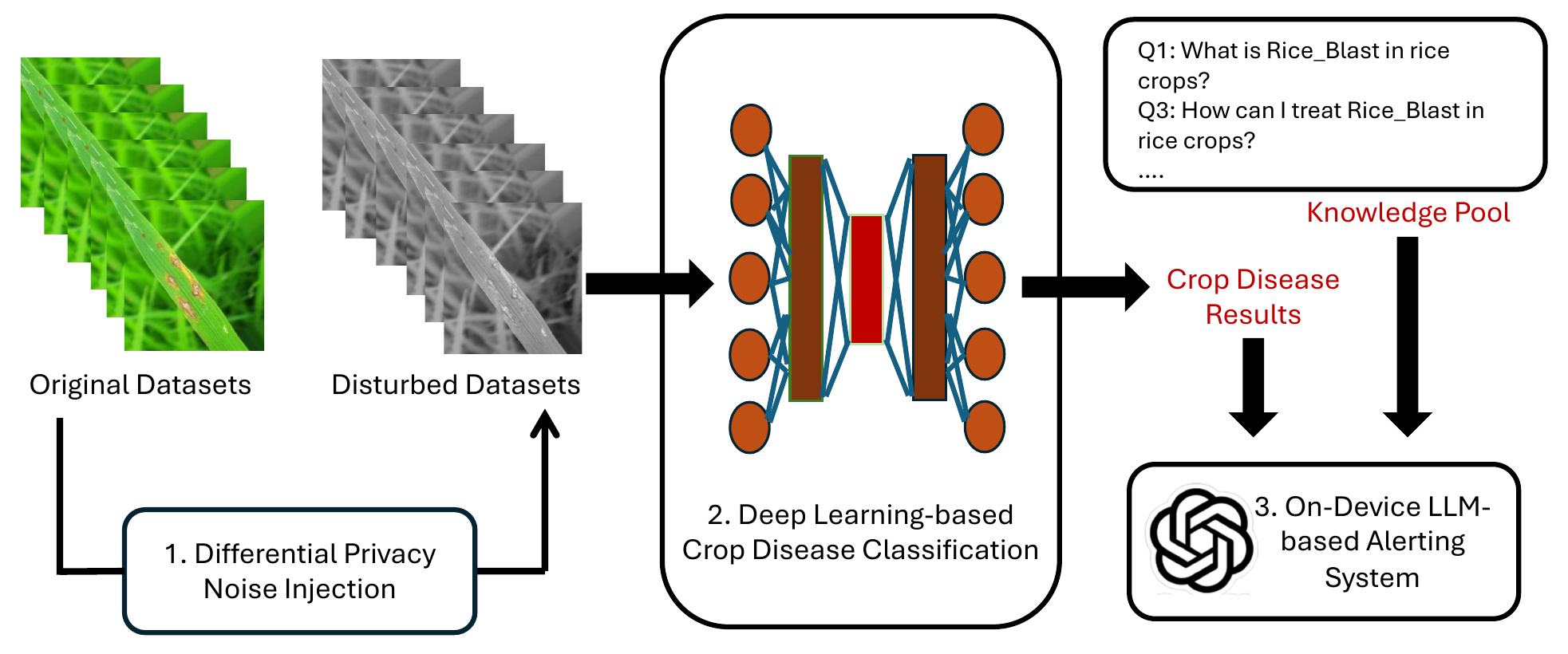}
  \centering
  \caption{AgriSentinel Framework}
  \label{fig:AgriSentinel}
\Description{fig:AgriSentinel}
\end{figure*}

We propose an AgriSentinel framework, illustrated in Fig.~\ref{fig:AgriSentinel}, as a privacy-enhanced on-device system for comprehensive crop disease alerting that seamlessly integrates differential privacy, deep learning, and LLM. 
The framework begins with a Gaussian noise-based differential privacy mechanism to safeguard sensitive crop data, introducing controlled noise to achieve a balance between privacy protection and the preservation of critical diagnostic features. 
The obfuscated data is subsequently processed by a lightweight deep learning model designed for efficient and accurate disease classification, leveraging advanced architectural components such as convolutional layers and ReLU activation to optimize computational efficiency and predictive accuracy. 
Building upon these predictive results, the framework incorporates a fine-tuned on-device LLM, enriched with a curated knowledge pool, to generate precise and actionable disease management strategies. 
To sum up, AgriSentinel offers farmers a lightweight and user-friendly tool to enhance crop health management while addressing key challenges in usability and data protection.

\subsection{Differential Privacy Noise Injection Mechanism}\label{subsec:DP}

We propose a differential privacy mechanism to protect crop data by employing Gaussian noise as an effective obfuscation technique. Gaussian noise introduces controlled random variations to the pixel values of an image, masking identifiable features while preserving the image’s utility for analysis. This approach strikes a balance between safeguarding data privacy and retaining the essential diagnostic information required for accurate disease pattern recognition. The process of applying Gaussian noise to an input image $x$ can be mathematically expressed as:
\begin{equation}
x_{\text{obf}} = x + N(0, \sigma^2),
\end{equation} where $I_{\text{obf}}$ is the resulting obfuscated image and \( Lap(0, \sigma^2) \) represents Gaussian noise with a mean of 0 and a tunable variance \(\sigma^2\).
The mean value of 0 ensures that the noise is centered around the original pixel value, maintaining an unbiased perturbation, while the variance \( \sigma^2 \) dictates the level of noise introduced. 
A higher variance results in more pronounced noise, providing stronger obfuscation and enhanced privacy protection but potentially impacting image clarity and the model's ability to interpret diagnostic features. Conversely, a lower variance applies subtler noise, which maintains more image detail but offers less privacy shielding.

The proposed differential privacy noise injection mechanism ensures that core diagnostic features essential for disease recognition remain intact. The strategic application of Gaussian noise enables the system to comply with data privacy standards while allowing the machine learning model to access sufficient visual information for accurate disease classification. The obfuscated images are suitable for processing by deep learning algorithms, which can learn robust features for pattern recognition even when some noise is present.
This approach is particularly advantageous for applications where image data needs to be shared across different systems or with third parties, maintaining data confidentiality and minimizing the risk of privacy breaches. The flexibility in tuning the variance \( \sigma^2 \) allows for a customizable level of obfuscation that can be adapted based on specific privacy requirements or regulations.

\subsection{Deep Learning-based Crop Disease Classification}\label{subsec:DLCDC}

After acquiring the obfuscated crop images dataset, we have developed a lightweight deep learning model tailored for crop disease classification. This model is structured as follows:
(1) Data Normalization: An essential preprocessing step, data normalization adjusts each pixel value by dividing by 255. This scales the image pixel values to a range of [0, 1], ensuring a uniform distribution of input values. Such normalization not only stabilizes the learning process but also accelerates convergence and improves the accuracy of the model by minimizing the impact of variations in image brightness and contrast.
(2) Convolutional and Pooling Layers: These layers are crucial for extracting spatial and local features from the images of rice diseases. The convolutional layers detect these features, while the pooling layers reduce the dimensionality of the data, which helps in handling overfitting.
(3) Dense and Activation Layers: Positioned after the convolutional and pooling layers, the dense layer, coupled with a ReLU activation layer in Eq.~\eqref{eq:dense_relu}, integrates the extracted features. This combination is pivotal for the complex decision-making required to accurately predict disease categories.
By streamlining the architecture with these elements, our model remains lightweight yet powerful enough to effectively classify various crop diseases based on visual cues.
\begin{equation}
\label{eq:dense_relu}
y = \text{ReLU}(W \cdot x_{obf} + b),
\end{equation} where \( x_{obf} \) is the input obfuscated crop dataset, \( W \) is the weight matrix learned by convolutional layer and pooling layers during training, \( b \) is the bias vector, and \text{ReLU} stands for the activation function applied to the output.

This proposed lightweight deep learning model is designed to classify crop images into various disease categories. Central to training this neural network for multi-class classification is the cross-entropy loss function, which measures the discrepancy between the true labels and the model’s predicted probabilities. Categorical cross-entropy effectively gauges how closely the predicted probability distribution matches the actual distribution, providing a clear pathway for the model to refine its predictions. The mathematical formulation of this loss function is presented in Eq.~\eqref{eq:cross_entropy}. For optimization, the model employs the Adam optimizer, known for its ability to adjust the learning rate dynamically throughout the training process. The training itself utilizes a fitting method, iteratively adjusting the model’s weights based on the training data to enhance performance and accuracy.
\begin{equation}
\label{eq:cross_entropy}
\mathcal{L} = -\sum_{c=1}^{C} y_{c} \log(\hat{y}_{c}),
\end{equation} where \( \mathcal{L} \) is the overall loss value that the model aims to minimize during training, \( C \) is the total number of classes, \( y_{c} \) is the true label for class \( c \), which takes a value of 1 if the input image belongs to class \( c \) and 0 otherwise, and \( \hat{y}_{c} = P(y = c \mid x) = \frac{e^{z_c}}{\sum_{j=1}^{C} e^{z_j}}\) is the predicted probability for class \( c \) produced by the softmax output of the model.

\subsection{On-Device LLM-based Alerting System}\label{subsec:LLM_system}

After successfully classifying crop diseases, our system provides precise diagnostic results. 
To enhance this functionality, we have established a knowledge pool consisting of question-and-answer pairs $K = (Q, A)$ related to crop disease management, serving as a foundational dataset for fine-tuning an on-device LLM-based alerting system $AgriLLM(\cdot)$. 
When a disease is detected, the model uses the diagnostic results as input, utilizing questions from the knowledge pool as prompts to generate relevant answers. 
This integration facilitates the creation of alerts and therapeutic suggestions tailored to the specific disease identified. 
The responses $R$ generated by the LLM offer concrete advice on managing the disease, providing actionable insights to end-users such as farmers or agronomists.
The process of the proposed on-device LLM-based alerting system can be formulated as below:
\begin{equation}
\label{eq:agri_alert}
R = AgriLLM(\hat{y},(Q,A);\theta),
\end{equation} where $\hat{y}$ is the predicted crop disease result, $\theta$ represents the parameters of the fine-tuned on-device LLM-based alerting system and $R$ should be generated as similar as the groundtruth answers $A$.
This empowers them with better decision-making tools and effective treatment strategies, optimizing crop health management.

\section{Experiment}\label{sec:experiment}

In this section, we outline the experimental setup and provide detailed results to demonstrate the effectiveness of our proposed crop disease alerting system in delivering accurate performance while ensuring data privacy protection.

\subsection{Experiment Settings}\label{subsec:exp_set}

We present the experimental setup across several key aspects: dataset selection, data preprocessing, model architecture, training methodology, LLM fine-tuning, and performance evaluation metrics.

\subsubsection{Dataset}\label{subsubsec:datasets}

\begin{figure*}[htbp]
\centering
\includegraphics[width=\textwidth]{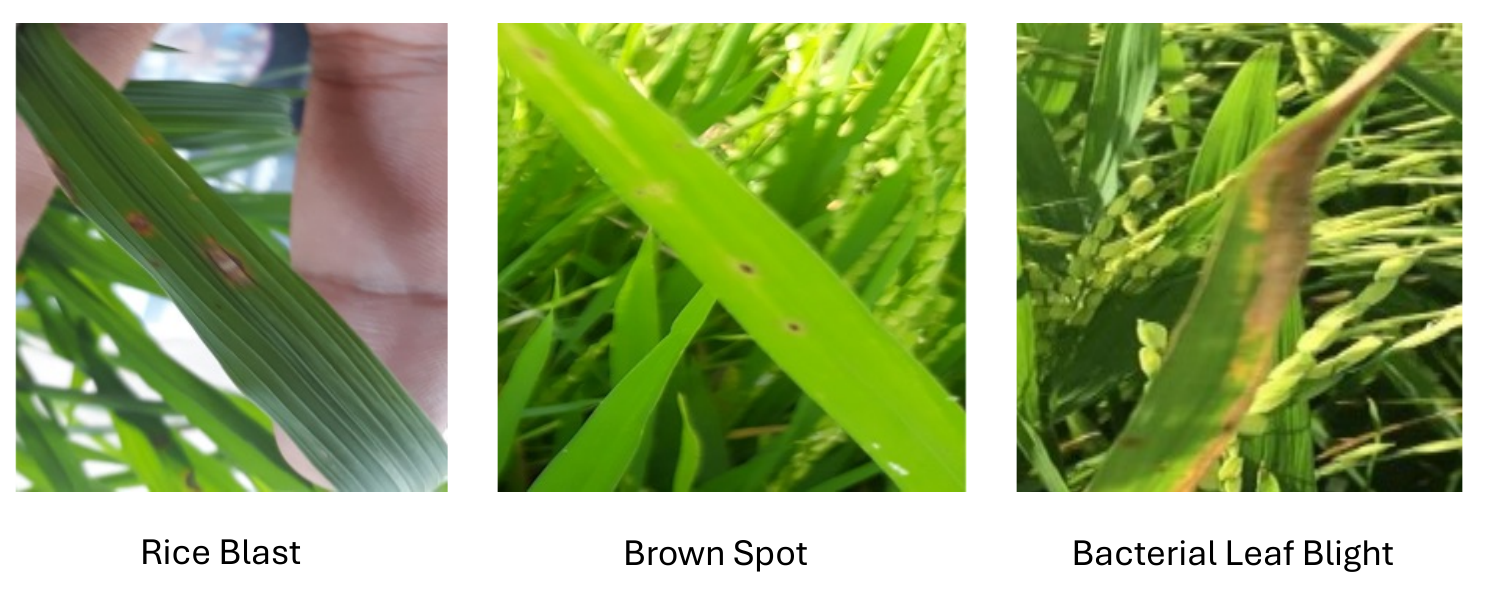}
\caption{Three Primary Categories of Rice Plant Diseases}
\label{fig:dataset}
\Description{fig:dataset}
\end{figure*}

In this paper, we use the dataset, including images of rice plants to discern numerous diseases that affect those rice crops~\cite{narmadha2022deep}.
Crop images in the collection are arranged into three primary categories of rice plant diseases in Fig.~\ref{fig:dataset}.
(1) \textbf{Rice Blast}: Oval shaped lesions appearing on leaves caused by the fungus Magnaporthe oryzae leads to yield reduction, which is typified by this disease.
(2) \textbf{Brown Spot}: This disease causes affects of photosynthesis and hence health of the plant caused by the fungus Bipolaris oryzae and appear as dark brown patches on leaves.
(3) \textbf{Bacterial Leaf Blight}: It is caused by a bacterium (Xanthomonas oryzae) that initially manifests as yellow streaks on the leaves, and from which the leaves eventually fall off and production is reduced.
There are totally 1,200 photos in the collection, split equally across the three classifications, and each one is annotated with the ailment it depicts. 
In order to facilitate effective access and retrieval throughout the training and evaluation phases, the dataset is arranged into folders, each labeled after the corresponding class.
The dataset was gathered from a variety of agricultural sources and field research, guaranteeing a representative sample of rice illnesses found in different geographic areas.

\begin{table}[htbp]
\caption{Different Levels of Privacy Loss}
\centering
\resizebox{\linewidth}{!}{\begin{tabular}{|c|c|c|c|}
\hline
\textbf{Obfuscation Level} & \textbf{Sensitivity (\(\Delta f\))} & \textbf{Noise Scale (\(\sigma\))} & \textbf{Privacy Loss (\(\epsilon\))} \\ \hline
Very High   & 0.1 & 1.5 & 0.06  \\ \hline
High        & 0.3 & 1.0& 0.30  \\ \hline
Medium      & 1.0 & 0.5 & 2.0  \\ \hline
Low     & 2.0 & 0.2 & 10.0  \\ \hline
Very Low     & 3.0 & 0.1 & 30.0  \\ \hline
\end{tabular}}
\label{tab:privacy_loss_evl}
\end{table}

\subsubsection{Data Preprocessing}\label{subsubsec:data_pre}

Data preparation is a crucial step before feeding images into the proposed lightweight deep learning-based crop disease classification model, ensuring the data is properly formatted for effective training. In this study, all images were resized to 224x224 pixels, a standard size required for CNN models to maintain consistent input dimensions with pre-trained architectures. Pixel values were normalized by dividing each by 255, scaling them to the [0, 1] range, which stabilizes the learning process, accelerates convergence, and improves model accuracy by reducing the impact of brightness and contrast variations. Labels for disease categories were encoded into numerical values, enabling the model to interpret categorical data effectively during training. The dataset was then split into training and testing sets, ensuring 80\% of the data was used for training and 20\% for testing to evaluate the model’s performance.

\subsubsection{Deep Learning Model Architecture}\label{subsubsec:nn_arch}

The proposed deep learning architecture is designed to effectively classify crop disease images. It begins with an input layer that accepts images of size $224 \times 224 \times 3$, corresponding to the standard RGB format. The first convolutional layer applies 32 filters of size $3 \times 3$ with ReLU activation, followed by a max pooling layer to reduce the spatial dimensions of the feature maps. This is succeeded by a second convolutional layer with 64 filters of the same size and ReLU activation, paired with another max pooling layer. The extracted 2D feature maps are then flattened into a 1D vector by the flattening layer for further processing. A dense layer with 128 units and ReLU activation follows, accompanied by a dropout layer to mitigate overfitting. Finally, the architecture employs a softmax output layer, which calculates class probabilities.

\begin{table}[ht]
\centering
\caption{Performance of Classifying Rice Blast with Different Privacy Loss}
\label{tab:riceblast_metrics}
\resizebox{\linewidth}{!}{\begin{tabular}{cccccc}
\toprule
\textbf{Obfuscation Level} & \textbf{Privacy Loss ($\epsilon$)} & \textbf{Accuracy} & \textbf{Precision} &  \textbf{Recall} &  \textbf{F1\_Score} \\
\midrule
 &      0 (Baseline) &    0.84 &   0.97 &    0.89 &      0.93 \\
Very High &      0.06 &    0.65 &   0.68 &    0.65 &      0.66 \\
High &      0.30 &    0.78 &   0.80 &    0.78 &      0.79 \\
Medium &      2.0 &     0.85 &   0.87 &    0.85 &      0.86 \\
Low &    10.0 &    0.88 &   0.91 &    0.89 &      0.90 \\
Very Low &     30.0 &    0.91 &   0.93 &    0.91 &      0.92 \\
\bottomrule
\end{tabular}}
\end{table}

\begin{table}[ht]
\centering
\caption{Performance of Classifying Brown Spot with Different Privacy Loss}
\label{tab:brownspot_metrics}
\resizebox{\linewidth}{!}{\begin{tabular}{cccccc}
\toprule
\textbf{Obfuscation Level} & \textbf{Privacy Loss ($\epsilon$)} & \textbf{Accuracy} & \textbf{Precision} &  \textbf{Recall} &  \textbf{F1\_Score} \\
\midrule
 &        0 (Baseline) &  0.87 &    0.79 & 0.81 & 0.80 \\
Very High &     0.06 &    0.62 &   0.65 &    0.62 &      0.63 \\
High &      0.30 &    0.76 &   0.78 &    0.76 &      0.77 \\
Medium &      2.0 &     0.83 &   0.85 &    0.83 &      0.84 \\
Low &      10.0 &    0.87 &   0.89 &    0.87 &      0.88 \\
Very Low &      30.0 &    0.90 &   0.92 &    0.90 &      0.91 \\
\bottomrule
\end{tabular}}
\end{table}

\begin{table}[ht]
\centering
\caption{Performance of Classifying Bacterial Leaf Blight with Different Privacy Loss}
\label{tab:bacterial_metrics}
\resizebox{\linewidth}{!}{\begin{tabular}{cccccc}
\toprule
\textbf{Obfuscation Level} & \textbf{Privacy Loss ($\epsilon$)} & \textbf{Accuracy} & \textbf{Precision} &  \textbf{Recall} &  \textbf{F1\_Score} \\
\midrule
&         0 (Baseline) &  0.87 &    0.45 & 0.86 & 0.59 \\
Very High &      0.06 &    0.60 &   0.63 &    0.60 &      0.61 \\
High &      0.30 &    0.72 &   0.75 &    0.72 &      0.73 \\
Medium &      2.0 &     0.81 &   0.83 &    0.81 &      0.82 \\
Low &      10.0 &    0.85 &   0.88 &    0.85 &      0.86 \\
Very Low &      30.0 &    0.89 &   0.91 &    0.89 &      0.90 \\
\bottomrule
\end{tabular}}
\end{table}

\begin{table*}[htbp]
\renewcommand{\arraystretch}{1.5} 
\caption{Performance of Our Proposed LLM-based Alerting System}
\resizebox{\textwidth}{!}{\begin{tabular}{|m{2.5cm}|m{7cm}|m{7cm}|}
\hline
\textbf{Prompt (Questions with Predicted Disease Result)} & \textbf{Fine-Tuned LLM's Answer} & \textbf{Groundtruth in Knowledge Pool} \\ \hline
How can I prevent \textbf{Brown Spot} rice crops in the future?  & There are several measures you can take to prevent brown spot in rice crops in the future: Plant disease-resistant rice varieties: Choose rice varieties that are resistant to brown spot to reduce the risk of infection. Rotate crops: {\color{red}Rotate rice with other crops to reduce the build-up of pathogens in the soil that cause brown spot. Practice good field management:} Implement proper irrigation and drainage practices to prevent waterlogging, which can create conditions conducive to brown spot development. Use clean seeds: Plant disease-free seeds to prevent the introduction of pathogens into your rice crop. Apply fungicides: If brown spot is a recurring problem in your area, consider applying fungicides to protect your rice crop from infection. Monitor crop regularly   &   To prevent brown spot in rice crops~\cite{macasero2024influence}, consider the following measures:
•	Plant disease-resistant varieties to reduce infection risk.
•	Rotate crops to minimize pathogen build-up in the soil.
•	{\color{red}Practice good field management with proper irrigation }and drainage to prevent water stress~\cite{crill1982role}.
•	Use clean seeds to avoid introducing pathogens
•	Apply fungicides if brown spot is common in your area.
•	Monitor crops regularly for early detection and intervention~\cite{nguefack2013effect}
 \\ \hline

What are the environmental conditions that favor rice crops \textbf{Brown Spot}? &  The environmental conditions that favor rice brownspot, caused by the fungus Bipolaris oryzae, include {\color{red}warm and humid weather}, as well as frequent rainfall or high levels of humidity. These conditions create a suitable environment for the pathogen to thrive and infect rice plants, particularly during the rainy season. Additionally, {\color{red}overcrowded and poorly ventilated fields} can also promote the spread of the disease by creating a suitable microclimate for the fungus to grow and spread. Inadequate drainage in the field can also contribute to the development of rice brownspot, as standing water provides a favorable environment for the fungus to infect the plants. & •	{\color{red}Warm and Humid Weather: Optimal for the growth of the Bipolaris oryzae fungus}~\cite{singh2012diseases}.
•	{\color{red}Frequent Rainfall or High Humidity}: Increases moisture levels, creating favorable conditions for the pathogen.
•	Overcrowded Fields: Reduces airflow, creating a microclimate that supports fungal growth~\cite{mo2012analysis}.
•	{\color{red}Poor Ventilation}: Contributes to moisture retention and a conducive environment for infection.
•	{\color{red}Inadequate Drainage}: Leads to standing water, further supporting fungal development and spread~\cite{fujisaka1994rice}.
 \\ \hline

Factors affecting the life cycle and rice \textbf{Brown Spot} epidemics?  & 1. Climate: Brown spot disease is more prevalent in areas with warm, humid climates, as these conditions are conducive to the growth and spread of the fungal pathogen responsible for the disease.
2.	{\color{red}Soil type}: Certain soil types may be more favorable for the development and spread of brown spot disease, as they provide the ideal conditions for the fungal pathogen to thrive.
3.	{\color{red}Rice variety}: Some rice varieties may be more susceptible to brown spot disease than others. Varieties with genetic resistance to the disease are less likely to be affected.
4.	{\color{red}Crop rotation}: Continuous planting of rice crops in the same field can increase the prevalence of brown spot disease. Rotating rice crops with other crops can help reduce the risk of infection.
5.	{\color{red}Management practices} & Factors affecting the life cycle and rice brown spot epidemics include:
1.	Climate: Warm, humid conditions support the growth and spread of the Bipolaris oryzae fungus, increasing the risk of epidemics~\cite{lirova2023reaction}.
2.	{\color{red}Soil Type}: Poorly drained or low-fertility soils can contribute to disease development.
3.	{\color{red}Rice Variety}: Susceptibility varies; varieties lacking genetic resistance are more prone to infection.
4.	{\color{red}Crop Rotation}: Continuous rice planting increases disease prevalence. Crop rotation with non-host crops helps disrupt the pathogen's cycle~\cite{berhan2021morphological}.
5.	{\color{red}Management Practices}: Poor field management, including suboptimal fertilization and improper irrigation, can create favorable conditions for the disease~\cite{singh2012cultural}.
\\ \hline
\end{tabular}}
\label{tab:LLM_perf}
\end{table*}

\begin{table*}[htbp]
\renewcommand{\arraystretch}{1.5} 
\caption{Performance of Our Proposed LLM-based Alerting System (Continued)}
\resizebox{\textwidth}{!}{\begin{tabular}{|m{2.5cm}|m{7cm}|m{7cm}|}
\hline
Can \textbf{Brown Spot} spread to other rice crops?  &  Yes, {\color{red}Brownspot, a fungal disease caused by Bipolaris oryzae}, can spread to other rice crops if proper measures are not taken to control its spread. The disease is commonly spread through wind, rain, and agricultural practices that facilitate the transmission of the fungal spores. To prevent the spread of Brownspot to other rice crops, it is important to implement effective disease management strategies such as {\color{red}crop rotation}, {\color{red}planting resistant varieties}, and using fungicides when necessary. Additionally, practicing good sanitation and proper field management can help minimize the spread of the disease.   &   {\color{red}Brownspot, a fungal disease caused by Bipolaris oryzae}, can spread to other rice crops if not properly managed~\cite{berhan2021morphological}. The fungus spreads through wind and rain, which carry the spores from infected plants to healthy ones. Agricultural practices, such as using contaminated equipment or not removing infected plant debris, can also facilitate the spread~\cite{parthasarathy2020diseases}. To prevent the spread of Brownspot, it is crucial to:
•	{\color{red}Implement crop rotation with non-host crops.}
•	Plant disease-resistant rice varieties.
•	{\color{red}Apply fungicides when necessary and according to local agricultural guidelines}
•	Practice good field sanitation by removing infected plant material.
•	Maintain proper field management, such as adequate spacing for airflow and effective irrigation practices~\cite{lirova2023reaction}.

\\ \hline
What chemicals should I use to treat rice \textbf{Brown Spot}? & 1. Fungicides such as {\color{red}azoxystrobin, flutolanil, or prothioconazole} can be used to treat rice Brownspot. These chemicals work by inhibiting the growth of the fungus that causes the disease.
2.	Insecticides such as {\color{red}chlorpyrifos or malathion} may also be used to control insect pests that can contribute to the spread of Brownspot.
3.	It is important to carefully follow the application instructions provided on the label of the chemical product to ensure safe and effective treatment of Brownspot in rice. Additionally, it is recommended to rotate between different fungicides with different modes of action to prevent the development of resistance in the fungal population.
  &   Recommended chemicals for treating rice brownspot include:
1.	Fungicides: Use fungicides such as {\color{red} azoxystrobin, flutolanil, or prothioconazole}, which are effective in controlling the growth of the {\color{red}Bipolaris oryzae} fungus.
2.	Application Practices:
Always follow the label instructions for safe and effective application.Rotate between fungicides with different modes of action to prevent resistance development in the fungal population~\cite{van2019evaluation}.

\\ \hline
How do I improve soil health to prevent rice \textbf{Brown Spot}?   & Improving soil health to prevent rice brown spot:
1.	{\color{red}Implement crop rotation}: Rotate rice with non-host crops to break the disease cycle and reduce pathogen build-up in the soil.
2.	{\color{red}Incorporate organic matter}: Add compost or organic matter to improve soil structure, fertility, and biological activity, which makes it less favorable for pathogens.
3.	{\color{red}Practice proper irrigation}: Avoid over-irrigation and ensure good drainage to prevent waterlogged conditions that can promote pathogen growth.
4.	Use disease-resistant varieties: Plant rice varieties that are resistant to brown spot to reduce the risk of infection
    &  1.	{\color{red}Crop rotation}: This is an effective strategy to disrupt the disease cycle and prevent pathogen build-up. Rotating rice with non-host crops reduces the pathogen load in the soil, making it less likely for diseases like brown spot to persist~\cite{mishra2013diseases}.
2.	{\color{red}Incorporate organic matter}: Adding compost or organic matter enriches the soil, improving its structure and microbial activity, which can suppress harmful pathogens and promote overall soil health.
3.	{\color{red}Proper irrigation practices}: Managing water levels to avoid waterlogging is crucial. Over-irrigation creates conditions that can foster pathogen growth. {\color{red}Ensuring well-drained}fields reduces the risk of brown spot and other diseases that thrive in {\color{red}excessive moisture}~\cite{kharayat2023diseases}.
4.	{\color{red}Disease-resistant varieties}: Planting resistant rice varieties is an effective way to minimize infection risk and reduce reliance on chemical treatments~\cite{barnwal2013review}.
\\ \hline

\end{tabular}}
\label{tab:LLM_perf_2}
\end{table*}

\begin{table*}[ht]
\centering
\captionsetup{justification=centering}
\caption{Illustrative Example for Crop Image with Rice Blast}
\label{tab:example_rice_blast}
\resizebox{\textwidth}{!}{
\begin{tabular}{|m{3cm}|m{12cm}|}
        \hline
        \multicolumn{2}{|c|}{%
            \includegraphics[width=0.4\textwidth]{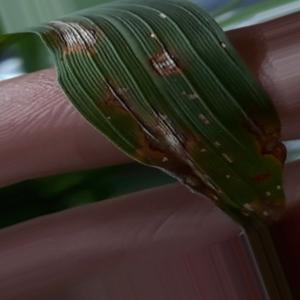} \hspace{1cm}%
            \includegraphics[width=0.4\textwidth]{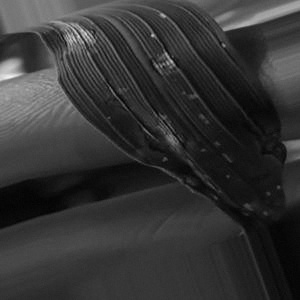}
        } \\
        \multicolumn{2}{|c|}{%
            (a) Original Crop Image with Rice Blast \hspace{1cm} (b) Privacy-Enhanced Crop Image with Rice Blast
        } \\
        
        \hline
        \textbf{Prompt with Disease Prediction Result} & Q: What chemicals should I use to treat \textbf{Rice Blast}?  \\

        \hline
        \textbf{Original GPT-2's Answer (Baseline)} & A: There are several chemicals that can be used to treat Blast, a fungal disease that affects a variety of plants. Some commonly used chemicals for treating Blast include: Fungicides: Fungicides such as azoxystrobin, tebuconazole, and propiconazole are effective in treating Blast by controlling the growth and spread of the fungal pathogen. These fungicides can be applied as foliar sprays or drenches to provide protection to the plants. \\

        \hline
        \textbf{Fine-tuned LLM's Answer} & A: The treatment of Blast disease in plants typically involves the use of fungicides. Some common chemical fungicides that are effective in treating Blast disease include: Azoxystrobin, Pyraclostrobin, Trifloxystrobin, Tebuconazole, and Propiconazole. It is important to carefully follow the instructions on the fungicide label and use personal protective equipment when applying these chemicals. Additionally, rotating between different fungicides with different modes \\

        \hline
    \end{tabular}%
    }
\end{table*}

\begin{table*}[ht]
\centering
\captionsetup{justification=centering}
\caption{Illustrative Example for Crop Image with Brown Spot}
\label{tab:example_brownspot}
\resizebox{\textwidth}{!}{
\begin{tabular}{|m{3cm}|m{12cm}|}
        \hline
        \multicolumn{2}{|c|}{%
            \includegraphics[width=0.4\textwidth]{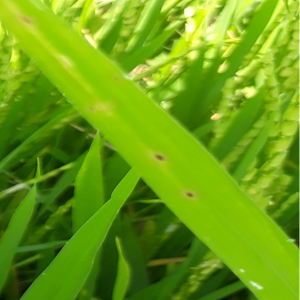} \hspace{1cm}%
            \includegraphics[width=0.4\textwidth]{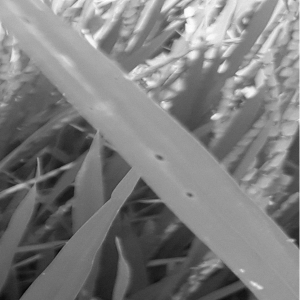}
        } \\
        \multicolumn{2}{|c|}{%
            (a) Original Crop Image with Brown Spot \hspace{1cm} (b) Privacy-Enhanced Crop Image with Brown Spot
        } \\
        
        \hline
        \textbf{Prompt with Disease Prediction Result} & Q: What chemicals should I use to treat \textbf{Brown Spot}?  \\

        \hline
        \textbf{Original GPT-2's Answer (Baseline)} & A: To treat Brownspot, you can use fungicides containing active ingredients such as azoxystrobin, propiconazole, or pyraclostrobin. Make sure to read and follow the manufacturer's instructions and safety precautions when using these chemicals. Additionally, good cultural practices such as proper irrigation and avoiding over-fertilization can help prevent the spread of Brownspot. \\

        \hline
        \textbf{Fine-tuned LLM's Answer} & A: There are several chemicals that can be used to effectively treat Brownspot, a fungal disease that commonly affects plants. Some options include: Chlorothalonil: This chemical is a broad-spectrum fungicide that can help control Brownspot and other fungal diseases on plants; Propiconazole: Another effective fungicide that can be used to treat Brownspot and other fungal diseases; Copper-based fungicides: These can also be effective in controlling Brownspot, especially when used preventatively. \\
              
        \hline
    \end{tabular}%
    }
\end{table*}

\begin{table*}[ht]
\centering
\captionsetup{justification=centering}
\caption{Illustrative Example for Crop Image with Bacterial Leaf Blight}
\label{tab:example_bacterial}
\resizebox{\textwidth}{!}{
\begin{tabular}{|m{3cm}|m{12cm}|}
        \hline
        \multicolumn{2}{|c|}{%
            \includegraphics[width=0.45\textwidth]{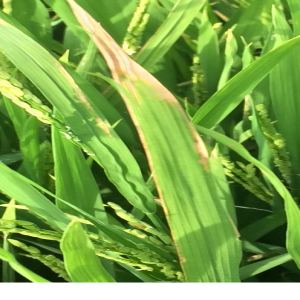} \hspace{1cm}%
            \includegraphics[width=0.45\textwidth]{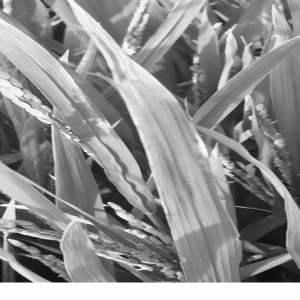}
        } \\
        \multicolumn{2}{|c|}{%
            (a) Original Crop Image with Bacterial Leaf Blight \hspace{1cm} (b) Privacy-Enhanced Crop Image with Bacterial Leaf Blight
        } \\
        
        \hline
        \textbf{Prompt with Disease Prediction Result} & Q: What chemicals should I use to treat \textbf{Bacterial Leaf Blight}?  \\

        \hline
        \textbf{Original GPT-2's Answer (Baseline)} & A: To treat bacterial leaf blight in plants, you can use chemicals such as copper-based fungicides, antibiotics like streptomycin, and bactericides like oxytetracycline. It is important to follow the instructions on the product label and apply the chemicals at the recommended rate and frequency to effectively control the disease. Additionally, rotating between different chemical treatments can help prevent the development of resistance in the bacteria.\\

        \hline
        \textbf{Fine-tuned LLM's Answer} & A: There are several chemicals that can be used to treat bacterial leaf blight in plants, including Copper-based fungicides: Copper sprays are commonly used to treat bacterial leaf blight as they have strong antibacterial properties and can help prevent the spread of the disease; Streptomycin: Streptomycin is an antibiotic that can be used to treat bacterial leaf blight, particularly in crops such as rice and corn; Oxytetracycline: Oxytetryield losses if left untreated. Bacterial leaf blight is favored by warm and humid conditions, making it a particular problem in regions with tropical or subtropical climates. \\
              
        \hline
    \end{tabular}%
    }
\end{table*}

\subsubsection{Deep Learning Model Training}\label{subsubsec:model_train}

To conceal sensitive details in the crop images, Gaussian noise will be added using our proposed differential privacy noise injection mechanism, ensuring that the data cannot be reconstructed or exploited by unauthorized users while still enabling the model to learn relevant features for disease detection.
The proposed deep learning-based crop disease classification model then will be trained with the obfuscated crop images.
The training will be performed for a definite batch size of 16 with the early stoppage scheme and learning rate reduction and the training data augmentation techniques were used.
The training for the model continued for several epochs until the resulting validation accuracy stopped increasing to the maximum.

\subsubsection{Integration with LLM for Alerts}\label{subsubsec:LLM_alert}

We utilize GPT-2 as the foundational LLM model and fine-tune it using our curated knowledge pool on crop disease management and the predicted crop disease results. This fine-tuned, on-device LLM-based alerting system provides farmers with timely crop disease alerts and tailored therapeutic suggestions.

\subsubsection{Performance Metrics}\label{subsubsec:performance_metrics}

We use accuracy, precision, recall and F1 score as the metrics for the deep learning performance comparison.
Besides, the metric Privacy Loss \( \epsilon \) is used to assess the effectiveness of a privacy-preserving framework in obfuscating sensitive information within images. 
In the context of rice disease data, privacy loss quantifies the extent to which identifiable information remains visible after obfuscation. 
The calculation of privacy loss is based on two primary factors: (1) Sensitivity of the input \( \Delta f \) and (2) Amount of Noise \( \sigma \). 
Sensitivity measures the maximum change in the output when a single individual's data is added or removed. In image obfuscation, sensitivity represents the degree of identifiable features retained in the image post-obfuscation.
Noise scale reflects the intensity of obfuscation applied to mask sensitive information. By adding Gaussian noise, the identifiable characteristics of the image are masked, reducing the recognizability of sensitive features.
To compute the overall privacy loss, both sensitivity and noise components are combined in the following manner: \(\epsilon = \frac{\Delta f}{\sigma}\).
This equation illustrates that privacy loss \( \epsilon \) is directly proportional to the sensitivity \( \Delta f \) and inversely proportional to the noise \( \sigma \) applied. Consequently, a higher sensitivity or lower noise results in a higher privacy loss, indicating more exposure of identifiable information. Conversely, lower sensitivity or higher noise diminishes privacy loss, thereby strengthening privacy protection.
We show five levels of privacy loss as examples in Table~\ref{tab:privacy_loss_evl}.

\subsection{Evaluation on Privacy-Enhanced Crop Disease Classification Model}\label{subsec:eval_privacy_performance}

We evaluate the proposed privacy-enhanced crop disease classification model using various performance metrics under different levels of privacy loss. As shown in Table~\ref{tab:riceblast_metrics}, the model without privacy protection serves as the baseline. Compared to the baseline, our model achieves adaptive privacy protection across different levels of privacy loss while maintaining accurate rice blast classification. When the obfuscation level is medium or higher, the model effectively balances data privacy protection and classification performance.
However, as the privacy protection level increases, a slight decline in classification performance is inevitable. Interestingly, we observe that the added noise for privacy protection enhances the robustness of the classification model, improving performance when the privacy protection level is below medium. Additionally, the comparison of classification performance for brown spot and bacterial leaf blight, presented in Table~\ref{tab:brownspot_metrics} and Table~\ref{tab:bacterial_metrics}, supports the same conclusion, demonstrating the model’s adaptability and effectiveness across different crop diseases.

\subsection{Evaluation on LLM-based Alerting System}\label{subsec:eval_LLM}

After obtaining the crop disease classification results, our LLM-based Alerting System is utilized to address questions related to the identified disease, providing farmers with actionable crop management suggestions. As shown in Table~\ref{tab:LLM_perf} and Table~\ref{tab:LLM_perf_2}, the fine-tuned LLM generates accurate responses closely aligned with expert-provided ground truth knowledge. This demonstrates that the LLM-based alerting system effectively delivers useful and timely suggestions to farmers, offering natural language explanations to support informed decision-making for managing crop diseases.

\subsection{Case Study of Our Proposed AgriSentinel System}\label{subsec:case_study}

We further demonstrate the effectiveness of our proposed AgriSentinel system through three illustrative examples. In Table~\ref{tab:example_rice_blast}, a crop image affected by rice blast is used as input to the AgriSentinel system. First, the image undergoes privacy enhancement using our differential privacy noise injection mechanism. Next, the lightweight deep learning-based crop disease classification model accurately identifies the crop disease in the image. Finally, questions related to the predicted crop disease are addressed by the LLM-based alerting model, providing farmers with actionable crop management suggestions. Additional examples in Table~\ref{tab:example_brownspot} and Table~\ref{tab:example_bacterial} highlight the system’s performance for brown spot and bacterial leaf blight diseases. These examples illustrate that AgriSentinel can automatically assist farmers by delivering effective crop disease management advice while safeguarding the privacy of crop images. Furthermore, the system’s lightweight design makes it suitable for deployment on mobile devices, ensuring ease of use and accessibility for farmers.

\section{Conclusion}\label{sec:conclusion}

In this paper, we propose AgriSentinel, a Privacy-Enhanced Embedded-LLM Crop Disease Alerting System. AgriSentinel integrates a differential privacy noise injection mechanism to protect crop image data, a lightweight deep learning model for accurate crop disease classification, and a fine-tuned LLM to provide farmers with strategic suggestions for crop disease alerting and management. Comprehensive experiments demonstrate that AgriSentinel effectively safeguards crop image privacy while delivering accurate and actionable recommendations for crop disease management. Additionally, the system is lightweight, easy to deploy, and designed to be farmer-friendly, enabling automated crop disease alerting and management, ultimately contributing to improved crop production.

\bibliographystyle{ACM-Reference-Format}
\bibliography{AgriSentinel}


\begin{thebibliography}{45}


\ifx \showCODEN    \undefined \def \showCODEN     #1{\unskip}     \fi
\ifx \showDOI      \undefined \def \showDOI       #1{#1}\fi
\ifx \showISBNx    \undefined \def \showISBNx     #1{\unskip}     \fi
\ifx \showISBNxiii \undefined \def \showISBNxiii  #1{\unskip}     \fi
\ifx \showISSN     \undefined \def \showISSN      #1{\unskip}     \fi
\ifx \showLCCN     \undefined \def \showLCCN      #1{\unskip}     \fi
\ifx \shownote     \undefined \def \shownote      #1{#1}          \fi
\ifx \showarticletitle \undefined \def \showarticletitle #1{#1}   \fi
\ifx \showURL      \undefined \def \showURL       {\relax}        \fi
\providecommand\bibfield[2]{#2}
\providecommand\bibinfo[2]{#2}
\providecommand\natexlab[1]{#1}
\providecommand\showeprint[2][]{arXiv:#2}

\bibitem[\protect\citeauthoryear{Aggarwal, Khullar, Goyal, and Prola}{Aggarwal et~al\mbox{.}}{2024}]%
        {aggarwal2024resource}
\bibfield{author}{\bibinfo{person}{Meenakshi Aggarwal}, \bibinfo{person}{Vikas Khullar}, \bibinfo{person}{Nitin Goyal}, {and} \bibinfo{person}{Thomas~Andr{\'e} Prola}.} \bibinfo{year}{2024}\natexlab{}.
\newblock \showarticletitle{Resource-efficient federated learning over IoAT for rice leaf disease classification}.
\newblock \bibinfo{journal}{\emph{Computers and Electronics in Agriculture}}  \bibinfo{volume}{221} (\bibinfo{year}{2024}), \bibinfo{pages}{109001}.
\newblock


\bibitem[\protect\citeauthoryear{Andreev, Kotsenko, Varlamov, Kim, and Goryachkin}{Andreev et~al\mbox{.}}{2024}]%
        {andreev2024text}
\bibfield{author}{\bibinfo{person}{Aleksandr Andreev}, \bibinfo{person}{Anton Kotsenko}, \bibinfo{person}{Oleg Varlamov}, \bibinfo{person}{Radmir Kim}, {and} \bibinfo{person}{Boris Goryachkin}.} \bibinfo{year}{2024}\natexlab{}.
\newblock \showarticletitle{Text processing using LLM for automatic creation of agricultural crops knowledge bases}. In \bibinfo{booktitle}{\emph{BIO Web of Conferences}}, Vol.~\bibinfo{volume}{130}. EDP Sciences, \bibinfo{pages}{01029}.
\newblock


\bibitem[\protect\citeauthoryear{Aouedi, Vu, Sacco, Nguyen, Piamrat, Marchetto, and Pham}{Aouedi et~al\mbox{.}}{2024}]%
        {aouedi2024survey}
\bibfield{author}{\bibinfo{person}{Ons Aouedi}, \bibinfo{person}{Thai-Hoc Vu}, \bibinfo{person}{Alessio Sacco}, \bibinfo{person}{Dinh~C Nguyen}, \bibinfo{person}{Kandaraj Piamrat}, \bibinfo{person}{Guido Marchetto}, {and} \bibinfo{person}{Quoc-Viet Pham}.} \bibinfo{year}{2024}\natexlab{}.
\newblock \showarticletitle{A survey on intelligent Internet of Things: Applications, security, privacy, and future directions}.
\newblock \bibinfo{journal}{\emph{IEEE Communications Surveys \& Tutorials}} (\bibinfo{year}{2024}).
\newblock


\bibitem[\protect\citeauthoryear{Banerjee, Das, and Mondal}{Banerjee et~al\mbox{.}}{2024}]%
        {banerjee2024study}
\bibfield{author}{\bibinfo{person}{Saikat Banerjee}, \bibinfo{person}{Soumitra Das}, {and} \bibinfo{person}{Abhoy~Chand Mondal}.} \bibinfo{year}{2024}\natexlab{}.
\newblock \showarticletitle{A Study of the Application Domain of a Large Language Models in the Agricultural Sector}.
\newblock \bibinfo{journal}{\emph{International Journal of Innovative Research in Computer Science \& Technology}} \bibinfo{volume}{12}, \bibinfo{number}{5} (\bibinfo{year}{2024}), \bibinfo{pages}{74--78}.
\newblock


\bibitem[\protect\citeauthoryear{Barnwal, Kotasthane, Magculia, Mukherjee, Savary, Sharma, Singh, Singh, Sparks, Variar, et~al\mbox{.}}{Barnwal et~al\mbox{.}}{2013}]%
        {barnwal2013review}
\bibfield{author}{\bibinfo{person}{Manoj~Kumar Barnwal}, \bibinfo{person}{A Kotasthane}, \bibinfo{person}{Nicole Magculia}, \bibinfo{person}{Prasun~K Mukherjee}, \bibinfo{person}{Serge Savary}, \bibinfo{person}{Anand~Kumar Sharma}, \bibinfo{person}{Hanwant~B Singh}, \bibinfo{person}{US Singh}, \bibinfo{person}{AH Sparks}, \bibinfo{person}{Mukund Variar}, {et~al\mbox{.}}} \bibinfo{year}{2013}\natexlab{}.
\newblock \showarticletitle{A review on crop losses, epidemiology and disease management of rice brown spot to identify research priorities and knowledge gaps}.
\newblock \bibinfo{journal}{\emph{European Journal of Plant Pathology}}  \bibinfo{volume}{136} (\bibinfo{year}{2013}), \bibinfo{pages}{443--457}.
\newblock


\bibitem[\protect\citeauthoryear{Berhan~Ejigu}{Berhan~Ejigu}{2021}]%
        {berhan2021morphological}
\bibfield{author}{\bibinfo{person}{Muluadam Berhan~Ejigu}.} \bibinfo{year}{2021}\natexlab{}.
\newblock \emph{\bibinfo{title}{Morphological characterization of bipolaris oryzae and screening of brown spot disease resistant lowland rice (oryza sativa l.) genotypes under rainfed condition at fogera plain, south gondar administration zone in amhara region, Ethiopia}}.
\newblock \bibinfo{thesistype}{Ph.D. Dissertation}. \bibinfo{school}{Bahir Dar University}.
\newblock


\bibitem[\protect\citeauthoryear{Chen, Li, Dong, Hao, Xu, Huang, and Liu}{Chen et~al\mbox{.}}{2020}]%
        {chen2020practical}
\bibfield{author}{\bibinfo{person}{Hanxiao Chen}, \bibinfo{person}{Hongwei Li}, \bibinfo{person}{Guishan Dong}, \bibinfo{person}{Meng Hao}, \bibinfo{person}{Guowen Xu}, \bibinfo{person}{Xiaoming Huang}, {and} \bibinfo{person}{Zhe Liu}.} \bibinfo{year}{2020}\natexlab{}.
\newblock \showarticletitle{Practical membership inference attack against collaborative inference in industrial IoT}.
\newblock \bibinfo{journal}{\emph{IEEE Transactions on Industrial Informatics}} \bibinfo{volume}{18}, \bibinfo{number}{1} (\bibinfo{year}{2020}), \bibinfo{pages}{477--487}.
\newblock


\bibitem[\protect\citeauthoryear{Chen and Li}{Chen and Li}{2024a}]%
        {chen2024octopusv2}
\bibfield{author}{\bibinfo{person}{Wei Chen} {and} \bibinfo{person}{Zhiyuan Li}.} \bibinfo{year}{2024}\natexlab{a}.
\newblock \showarticletitle{Octopus v2: On-device language model for super agent}.
\newblock \bibinfo{journal}{\emph{arXiv preprint arXiv:2404.01744}} (\bibinfo{year}{2024}).
\newblock


\bibitem[\protect\citeauthoryear{Chen and Li}{Chen and Li}{2024b}]%
        {chen2024octopusv3}
\bibfield{author}{\bibinfo{person}{Wei Chen} {and} \bibinfo{person}{Zhiyuan Li}.} \bibinfo{year}{2024}\natexlab{b}.
\newblock \showarticletitle{Octopus v3: Technical report for on-device sub-billion multimodal ai agent}.
\newblock \bibinfo{journal}{\emph{arXiv preprint arXiv:2404.11459}} (\bibinfo{year}{2024}).
\newblock


\bibitem[\protect\citeauthoryear{Chen and Li}{Chen and Li}{2024c}]%
        {chen2024octopusv4}
\bibfield{author}{\bibinfo{person}{Wei Chen} {and} \bibinfo{person}{Zhiyuan Li}.} \bibinfo{year}{2024}\natexlab{c}.
\newblock \showarticletitle{Octopus v4: Graph of language models}.
\newblock \bibinfo{journal}{\emph{arXiv preprint arXiv:2404.19296}} (\bibinfo{year}{2024}).
\newblock


\bibitem[\protect\citeauthoryear{Chen, Li, Guo, and Shen}{Chen et~al\mbox{.}}{2024}]%
        {chen2024octo}
\bibfield{author}{\bibinfo{person}{Wei Chen}, \bibinfo{person}{Zhiyuan Li}, \bibinfo{person}{Zhen Guo}, {and} \bibinfo{person}{Yikang Shen}.} \bibinfo{year}{2024}\natexlab{}.
\newblock \showarticletitle{Octo-planner: On-device language model for planner-action agents}.
\newblock \bibinfo{journal}{\emph{arXiv preprint arXiv:2406.18082}} (\bibinfo{year}{2024}).
\newblock


\bibitem[\protect\citeauthoryear{Chillakuru, Divya, and Ananthajothi}{Chillakuru et~al\mbox{.}}{2022}]%
        {chillakuru2022enhanced}
\bibfield{author}{\bibinfo{person}{Prameeladevi Chillakuru}, \bibinfo{person}{D Divya}, {and} \bibinfo{person}{K Ananthajothi}.} \bibinfo{year}{2022}\natexlab{}.
\newblock \showarticletitle{Enhanced segmentation with optimized nine-layered CNN-based classification of leaf diseases: an automatic approach for plant disease diagnosis}.
\newblock \bibinfo{journal}{\emph{cybernetics and systems}} (\bibinfo{year}{2022}), \bibinfo{pages}{1--36}.
\newblock


\bibitem[\protect\citeauthoryear{Crill, Nuque, Estrada, and Bandong}{Crill et~al\mbox{.}}{1982}]%
        {crill1982role}
\bibfield{author}{\bibinfo{person}{Pat Crill}, \bibinfo{person}{FL Nuque}, \bibinfo{person}{BA Estrada}, {and} \bibinfo{person}{JM Bandong}.} \bibinfo{year}{1982}\natexlab{}.
\newblock \showarticletitle{The role of varietal resistance in disease management}.
\newblock \bibinfo{journal}{\emph{Evolution of the Gene Rotation Concept for Rice Blast Control}} (\bibinfo{year}{1982}), \bibinfo{pages}{103--121}.
\newblock


\bibitem[\protect\citeauthoryear{El~Mestari, Lenzini, and Demirci}{El~Mestari et~al\mbox{.}}{2024}]%
        {el2024preserving}
\bibfield{author}{\bibinfo{person}{Soumia~Zohra El~Mestari}, \bibinfo{person}{Gabriele Lenzini}, {and} \bibinfo{person}{Huseyin Demirci}.} \bibinfo{year}{2024}\natexlab{}.
\newblock \showarticletitle{Preserving data privacy in machine learning systems}.
\newblock \bibinfo{journal}{\emph{Computers \& Security}}  \bibinfo{volume}{137} (\bibinfo{year}{2024}), \bibinfo{pages}{103605}.
\newblock


\bibitem[\protect\citeauthoryear{Fujisaka, Harrington, and Hobbs}{Fujisaka et~al\mbox{.}}{1994}]%
        {fujisaka1994rice}
\bibfield{author}{\bibinfo{person}{Sam Fujisaka}, \bibinfo{person}{Larry Harrington}, {and} \bibinfo{person}{Peter Hobbs}.} \bibinfo{year}{1994}\natexlab{}.
\newblock \showarticletitle{Rice-wheat in South Asia: systems and long-term priorities established through diagnostic research}.
\newblock \bibinfo{journal}{\emph{Agricultural Systems}} \bibinfo{volume}{46}, \bibinfo{number}{2} (\bibinfo{year}{1994}), \bibinfo{pages}{169--187}.
\newblock


\bibitem[\protect\citeauthoryear{Gowda}{Gowda}{2024}]%
        {gowda2024secure}
\bibfield{author}{\bibinfo{person}{Sahana~D Gowda}.} \bibinfo{year}{2024}\natexlab{}.
\newblock \showarticletitle{Secure Multiparty Computation: Protocols, Collaborative Data Processing, and Real-World Applications in Industry}.
\newblock In \bibinfo{booktitle}{\emph{Cloud Security}}. \bibinfo{publisher}{Chapman and Hall/CRC}, \bibinfo{pages}{160--182}.
\newblock


\bibitem[\protect\citeauthoryear{Gupta, Abdelsalam, Khorsandroo, and Mittal}{Gupta et~al\mbox{.}}{2020}]%
        {gupta2020security}
\bibfield{author}{\bibinfo{person}{Maanak Gupta}, \bibinfo{person}{Mahmoud Abdelsalam}, \bibinfo{person}{Sajad Khorsandroo}, {and} \bibinfo{person}{Sudip Mittal}.} \bibinfo{year}{2020}\natexlab{}.
\newblock \showarticletitle{Security and privacy in smart farming: Challenges and opportunities}.
\newblock \bibinfo{journal}{\emph{IEEE access}}  \bibinfo{volume}{8} (\bibinfo{year}{2020}), \bibinfo{pages}{34564--34584}.
\newblock


\bibitem[\protect\citeauthoryear{Jalalzai, Kadoche, Leluc, and Plassier}{Jalalzai et~al\mbox{.}}{2022}]%
        {jalalzai2022membership}
\bibfield{author}{\bibinfo{person}{Hamid Jalalzai}, \bibinfo{person}{Elie Kadoche}, \bibinfo{person}{R{\'e}mi Leluc}, {and} \bibinfo{person}{Vincent Plassier}.} \bibinfo{year}{2022}\natexlab{}.
\newblock \showarticletitle{Membership Inference Attacks via Adversarial Examples}.
\newblock \bibinfo{journal}{\emph{arXiv preprint arXiv:2207.13572}} (\bibinfo{year}{2022}).
\newblock


\bibitem[\protect\citeauthoryear{Javaid, Haleem, Singh, and Suman}{Javaid et~al\mbox{.}}{2022}]%
        {javaid2022enhancing}
\bibfield{author}{\bibinfo{person}{Mohd Javaid}, \bibinfo{person}{Abid Haleem}, \bibinfo{person}{Ravi~Pratap Singh}, {and} \bibinfo{person}{Rajiv Suman}.} \bibinfo{year}{2022}\natexlab{}.
\newblock \showarticletitle{Enhancing smart farming through the applications of Agriculture 4.0 technologies}.
\newblock \bibinfo{journal}{\emph{International Journal of Intelligent Networks}}  \bibinfo{volume}{3} (\bibinfo{year}{2022}), \bibinfo{pages}{150--164}.
\newblock


\bibitem[\protect\citeauthoryear{Jimenez, Dorado, Cock, Prager, Delerce, Grillon, Andrade~Bejarano, Benavides, and Jarvis}{Jimenez et~al\mbox{.}}{2016}]%
        {jimenez2016observation}
\bibfield{author}{\bibinfo{person}{Daniel Jimenez}, \bibinfo{person}{Hugo Dorado}, \bibinfo{person}{James Cock}, \bibinfo{person}{Steven~D Prager}, \bibinfo{person}{Sylvain Delerce}, \bibinfo{person}{Alexandre Grillon}, \bibinfo{person}{Mercedes Andrade~Bejarano}, \bibinfo{person}{Hector Benavides}, {and} \bibinfo{person}{Andy Jarvis}.} \bibinfo{year}{2016}\natexlab{}.
\newblock \showarticletitle{From observation to information: data-driven understanding of on farm yield variation}.
\newblock \bibinfo{journal}{\emph{PloS one}} \bibinfo{volume}{11}, \bibinfo{number}{3} (\bibinfo{year}{2016}), \bibinfo{pages}{e0150015}.
\newblock


\bibitem[\protect\citeauthoryear{KHARAYAT}{KHARAYAT}{2023}]%
        {kharayat2023diseases}
\bibfield{author}{\bibinfo{person}{BHUPENDRA~SINGH KHARAYAT}.} \bibinfo{year}{2023}\natexlab{}.
\newblock \bibinfo{booktitle}{\emph{DISEASES OF FIELD AND HORTICULTURAL CROPS AND THEIR MANAGEMENT VOLUME--II}}.
\newblock \bibinfo{publisher}{PHI Learning Pvt. Ltd.}
\newblock


\bibitem[\protect\citeauthoryear{Liang and Shah}{Liang and Shah}{2023}]%
        {liang2023iot}
\bibfield{author}{\bibinfo{person}{Chen Liang} {and} \bibinfo{person}{Tufail Shah}.} \bibinfo{year}{2023}\natexlab{}.
\newblock \showarticletitle{IoT in agriculture: The future of precision monitoring and data-driven farming}.
\newblock \bibinfo{journal}{\emph{Eigenpub Review of Science and Technology}} \bibinfo{volume}{7}, \bibinfo{number}{1} (\bibinfo{year}{2023}), \bibinfo{pages}{85--104}.
\newblock


\bibitem[\protect\citeauthoryear{Lirova}{Lirova}{2023}]%
        {lirova2023reaction}
\bibfield{author}{\bibinfo{person}{Simon~V Lirova}.} \bibinfo{year}{2023}\natexlab{}.
\newblock \emph{\bibinfo{title}{Reaction of Rice Germplasm to Blast and Brownspot Under Natural and Controlled Environment in Mwea, Kirinyaga County, Kenya}}.
\newblock \bibinfo{thesistype}{Ph.D. Dissertation}. \bibinfo{school}{University of Nairobi}.
\newblock


\bibitem[\protect\citeauthoryear{Macasero, Castilla, Pangga, Marquez, Martin, Duque, and Laborte}{Macasero et~al\mbox{.}}{2024}]%
        {macasero2024influence}
\bibfield{author}{\bibinfo{person}{John Bethany~M Macasero}, \bibinfo{person}{Nancy~P Castilla}, \bibinfo{person}{Ireneo~B Pangga}, \bibinfo{person}{Leonardo~V Marquez}, \bibinfo{person}{Edwin~C Martin}, \bibinfo{person}{Ulysses~G Duque}, {and} \bibinfo{person}{Alice~G Laborte}.} \bibinfo{year}{2024}\natexlab{}.
\newblock \showarticletitle{Influence of production situation on the incidence of brown spot of rice (Oryza sativa) caused by Bipolaris oryzae in the Philippines}.
\newblock \bibinfo{journal}{\emph{Plant Pathology}} \bibinfo{volume}{73}, \bibinfo{number}{2} (\bibinfo{year}{2024}), \bibinfo{pages}{390--403}.
\newblock


\bibitem[\protect\citeauthoryear{Mehta, Kukreja, and Gupta}{Mehta et~al\mbox{.}}{2023}]%
        {mehta2023revolutionizing}
\bibfield{author}{\bibinfo{person}{Shiva Mehta}, \bibinfo{person}{Vinay Kukreja}, {and} \bibinfo{person}{Amit Gupta}.} \bibinfo{year}{2023}\natexlab{}.
\newblock \showarticletitle{Revolutionizing Maize Disease Management with Federated Learning CNNs: A Decentralized and Privacy-Sensitive Approach}. In \bibinfo{booktitle}{\emph{2023 4th International Conference for Emerging Technology (INCET)}}. IEEE, \bibinfo{pages}{1--6}.
\newblock


\bibitem[\protect\citeauthoryear{Mishra, Deepa, et~al\mbox{.}}{Mishra et~al\mbox{.}}{2021}]%
        {mishra2021automation}
\bibfield{author}{\bibinfo{person}{Dipali Mishra}, \bibinfo{person}{D Deepa}, {et~al\mbox{.}}} \bibinfo{year}{2021}\natexlab{}.
\newblock \showarticletitle{Automation and integration of growth monitoring in plants (with disease prediction) and crop prediction}.
\newblock \bibinfo{journal}{\emph{Materials Today: Proceedings}}  \bibinfo{volume}{43} (\bibinfo{year}{2021}), \bibinfo{pages}{3922--3927}.
\newblock


\bibitem[\protect\citeauthoryear{Mishra}{Mishra}{2013}]%
        {mishra2013diseases}
\bibfield{author}{\bibinfo{person}{RK Mishra}.} \bibinfo{year}{2013}\natexlab{}.
\newblock \bibinfo{booktitle}{\emph{Diseases Of Vegetable Crops And Their Integrated Management: A Colour Handbook}}.
\newblock \bibinfo{publisher}{New India Publishing Agency}.
\newblock


\bibitem[\protect\citeauthoryear{Mo, Chen, and Yu}{Mo et~al\mbox{.}}{2012}]%
        {mo2012analysis}
\bibfield{author}{\bibinfo{person}{Jianguo Mo}, \bibinfo{person}{Qingyuan Chen}, {and} \bibinfo{person}{Fei Yu}.} \bibinfo{year}{2012}\natexlab{}.
\newblock \showarticletitle{Analysis of meteorological condition for infection of the pathogen of tobacco brown spot (Alternaria alternate)}.
\newblock \bibinfo{journal}{\emph{Plant Diseases and Pests}} \bibinfo{volume}{3}, \bibinfo{number}{2} (\bibinfo{year}{2012}), \bibinfo{pages}{24}.
\newblock


\bibitem[\protect\citeauthoryear{Narmadha, Sengottaiyan, and Kavitha}{Narmadha et~al\mbox{.}}{2022}]%
        {narmadha2022deep}
\bibfield{author}{\bibinfo{person}{RP Narmadha}, \bibinfo{person}{N Sengottaiyan}, {and} \bibinfo{person}{RJ Kavitha}.} \bibinfo{year}{2022}\natexlab{}.
\newblock \showarticletitle{Deep transfer learning based rice plant disease detection model.}
\newblock \bibinfo{journal}{\emph{Intelligent Automation \& Soft Computing}} \bibinfo{volume}{31}, \bibinfo{number}{2} (\bibinfo{year}{2022}).
\newblock


\bibitem[\protect\citeauthoryear{Nettleton, Katsantonis, Kalaitzidis, Sarafijanovic-Djukic, Puigdollers, and Confalonieri}{Nettleton et~al\mbox{.}}{2019}]%
        {nettleton2019predicting}
\bibfield{author}{\bibinfo{person}{David~F Nettleton}, \bibinfo{person}{Dimitrios Katsantonis}, \bibinfo{person}{Argyris Kalaitzidis}, \bibinfo{person}{Natasa Sarafijanovic-Djukic}, \bibinfo{person}{Pau Puigdollers}, {and} \bibinfo{person}{Roberto Confalonieri}.} \bibinfo{year}{2019}\natexlab{}.
\newblock \showarticletitle{Predicting rice blast disease: machine learning versus process-based models}.
\newblock \bibinfo{journal}{\emph{BMC bioinformatics}}  \bibinfo{volume}{20} (\bibinfo{year}{2019}), \bibinfo{pages}{1--16}.
\newblock


\bibitem[\protect\citeauthoryear{Nguefack, Wulff, Lekagne~Dongmo, Fouelefack, Fotio, Mbo, and Torp}{Nguefack et~al\mbox{.}}{2013}]%
        {nguefack2013effect}
\bibfield{author}{\bibinfo{person}{Julienne Nguefack}, \bibinfo{person}{G~Ednar Wulff}, \bibinfo{person}{J~Blaise Lekagne~Dongmo}, \bibinfo{person}{F~Romain Fouelefack}, \bibinfo{person}{Daniel Fotio}, \bibinfo{person}{Joseph Mbo}, {and} \bibinfo{person}{Jan Torp}.} \bibinfo{year}{2013}\natexlab{}.
\newblock \showarticletitle{Effect of plant extracts and an essential oil on the control of brown spot disease, tillering, number of panicles and yield increase in rice}.
\newblock \bibinfo{journal}{\emph{European Journal of Plant Pathology}}  \bibinfo{volume}{137} (\bibinfo{year}{2013}), \bibinfo{pages}{871--882}.
\newblock


\bibitem[\protect\citeauthoryear{Nong and Tarek}{Nong and Tarek}{2023}]%
        {nong2023surveillance}
\bibfield{author}{\bibinfo{person}{Nguyen~Bang Nong} {and} \bibinfo{person}{Mohammad Tarek}.} \bibinfo{year}{2023}\natexlab{}.
\newblock \showarticletitle{Surveillance Approaches to Intrusion Detection, Crop Health, and Disease Prevention in Agriculture}.
\newblock \bibinfo{journal}{\emph{Quarterly Journal of Emerging Technologies and Innovations}} \bibinfo{volume}{8}, \bibinfo{number}{3} (\bibinfo{year}{2023}), \bibinfo{pages}{1--17}.
\newblock


\bibitem[\protect\citeauthoryear{Parthasarathy, Thiribhuvanamala, and Prabakar}{Parthasarathy et~al\mbox{.}}{2020}]%
        {parthasarathy2020diseases}
\bibfield{author}{\bibinfo{person}{S Parthasarathy}, \bibinfo{person}{G Thiribhuvanamala}, {and} \bibinfo{person}{Kandasamy Prabakar}.} \bibinfo{year}{2020}\natexlab{}.
\newblock \bibinfo{booktitle}{\emph{Diseases of field crops and their management}}.
\newblock \bibinfo{publisher}{CRC Press}.
\newblock


\bibitem[\protect\citeauthoryear{Rahman, Fan, Zhang, and Chen}{Rahman et~al\mbox{.}}{2021}]%
        {rahman2021comparative}
\bibfield{author}{\bibinfo{person}{Mohammad Fatin~Fatihur Rahman}, \bibinfo{person}{Shurui Fan}, \bibinfo{person}{Yan Zhang}, {and} \bibinfo{person}{Lei Chen}.} \bibinfo{year}{2021}\natexlab{}.
\newblock \showarticletitle{A comparative study on application of unmanned aerial vehicle systems in agriculture}.
\newblock \bibinfo{journal}{\emph{Agriculture}} \bibinfo{volume}{11}, \bibinfo{number}{1} (\bibinfo{year}{2021}), \bibinfo{pages}{22}.
\newblock


\bibitem[\protect\citeauthoryear{Ristaino, Anderson, Bebber, Brauman, Cunniffe, Fedoroff, Finegold, Garrett, Gilligan, Jones, et~al\mbox{.}}{Ristaino et~al\mbox{.}}{2021}]%
        {ristaino2021persistent}
\bibfield{author}{\bibinfo{person}{Jean~B Ristaino}, \bibinfo{person}{Pamela~K Anderson}, \bibinfo{person}{Daniel~P Bebber}, \bibinfo{person}{Kate~A Brauman}, \bibinfo{person}{Nik~J Cunniffe}, \bibinfo{person}{Nina~V Fedoroff}, \bibinfo{person}{Cambria Finegold}, \bibinfo{person}{Karen~A Garrett}, \bibinfo{person}{Christopher~A Gilligan}, \bibinfo{person}{Christopher~M Jones}, {et~al\mbox{.}}} \bibinfo{year}{2021}\natexlab{}.
\newblock \showarticletitle{The persistent threat of emerging plant disease pandemics to global food security}.
\newblock \bibinfo{journal}{\emph{Proceedings of the National Academy of Sciences}} \bibinfo{volume}{118}, \bibinfo{number}{23} (\bibinfo{year}{2021}), \bibinfo{pages}{e2022239118}.
\newblock


\bibitem[\protect\citeauthoryear{Sapkota, Qureshi, Hassan, Shutske, Shoman, Sajjad, Dharejo, Paudel, Li, Meng, et~al\mbox{.}}{Sapkota et~al\mbox{.}}{2024}]%
        {sapkota2024multi}
\bibfield{author}{\bibinfo{person}{Ranjan Sapkota}, \bibinfo{person}{Rizwan Qureshi}, \bibinfo{person}{Syed~Zohaib Hassan}, \bibinfo{person}{John Shutske}, \bibinfo{person}{Maged Shoman}, \bibinfo{person}{Muhammad Sajjad}, \bibinfo{person}{Fayaz~Ali Dharejo}, \bibinfo{person}{Achyut Paudel}, \bibinfo{person}{Jiajia Li}, \bibinfo{person}{Zhichao Meng}, {et~al\mbox{.}}} \bibinfo{year}{2024}\natexlab{}.
\newblock \showarticletitle{Multi-Modal LLMs in Agriculture: A Comprehensive Review}.
\newblock \bibinfo{journal}{\emph{10.36227/techrxiv. 172651082.24507804/v1}} (\bibinfo{year}{2024}).
\newblock


\bibitem[\protect\citeauthoryear{Sharma et~al\mbox{.}}{Sharma et~al\mbox{.}}{2024}]%
        {sharma2024revolutionizing}
\bibfield{author}{\bibinfo{person}{Richa Sharma} {et~al\mbox{.}}} \bibinfo{year}{2024}\natexlab{}.
\newblock \showarticletitle{Revolutionizing Rice Agriculture: A Machine Learning Approach to Fungal Disease Management for Economic Sustainability}. In \bibinfo{booktitle}{\emph{2024 International Conference on Communication, Computer Sciences and Engineering (IC3SE)}}. IEEE, \bibinfo{pages}{798--805}.
\newblock


\bibitem[\protect\citeauthoryear{Silva, Nunes, Estev{\~a}o, Aski, and Chandra}{Silva et~al\mbox{.}}{2023}]%
        {silva2023gpt}
\bibfield{author}{\bibinfo{person}{Bruno Silva}, \bibinfo{person}{Leonardo Nunes}, \bibinfo{person}{Roberto Estev{\~a}o}, \bibinfo{person}{Vijay Aski}, {and} \bibinfo{person}{Ranveer Chandra}.} \bibinfo{year}{2023}\natexlab{}.
\newblock \showarticletitle{GPT-4 as an agronomist assistant? Answering agriculture exams using large language models}.
\newblock \bibinfo{journal}{\emph{arXiv preprint arXiv:2310.06225}} (\bibinfo{year}{2023}).
\newblock


\bibitem[\protect\citeauthoryear{Singh and Chawla}{Singh and Chawla}{2012}]%
        {singh2012cultural}
\bibfield{author}{\bibinfo{person}{Vaibhav~K Singh} {and} \bibinfo{person}{Shipli Chawla}.} \bibinfo{year}{2012}\natexlab{}.
\newblock \showarticletitle{Cultural Practices: An Ecofriendly Innovative Approach in Plant Disease Management}.
\newblock \bibinfo{journal}{\emph{International Book Publishers and Distributers, Dehradun--248}}  \bibinfo{volume}{1} (\bibinfo{year}{2012}), \bibinfo{pages}{01--20}.
\newblock


\bibitem[\protect\citeauthoryear{Singh, Singh, and Kumar}{Singh et~al\mbox{.}}{2012}]%
        {singh2012diseases}
\bibfield{author}{\bibinfo{person}{Vaibhav~K Singh}, \bibinfo{person}{Yogendra Singh}, {and} \bibinfo{person}{Prabhat Kumar}.} \bibinfo{year}{2012}\natexlab{}.
\newblock \showarticletitle{Diseases of ornamental plants and their management}.
\newblock \bibinfo{journal}{\emph{Eco-friendly innovative approaches in plant disease management}} (\bibinfo{year}{2012}), \bibinfo{pages}{543--572}.
\newblock


\bibitem[\protect\citeauthoryear{Sykuta}{Sykuta}{2016}]%
        {sykuta2016big}
\bibfield{author}{\bibinfo{person}{Michael~E Sykuta}.} \bibinfo{year}{2016}\natexlab{}.
\newblock \showarticletitle{Big data in agriculture: property rights, privacy and competition in ag data services}.
\newblock \bibinfo{journal}{\emph{International Food and Agribusiness Management Review}}  \bibinfo{volume}{19} (\bibinfo{year}{2016}), \bibinfo{pages}{57--74}.
\newblock


\bibitem[\protect\citeauthoryear{Tantalaki, Souravlas, and Roumeliotis}{Tantalaki et~al\mbox{.}}{2019}]%
        {tantalaki2019data}
\bibfield{author}{\bibinfo{person}{Nicoleta Tantalaki}, \bibinfo{person}{Stavros Souravlas}, {and} \bibinfo{person}{Manos Roumeliotis}.} \bibinfo{year}{2019}\natexlab{}.
\newblock \showarticletitle{Data-driven decision making in precision agriculture: The rise of big data in agricultural systems}.
\newblock \bibinfo{journal}{\emph{Journal of agricultural \& food information}} \bibinfo{volume}{20}, \bibinfo{number}{4} (\bibinfo{year}{2019}), \bibinfo{pages}{344--380}.
\newblock


\bibitem[\protect\citeauthoryear{Tzachor, Devare, Richards, Pypers, Ghosh, Koo, Johal, and King}{Tzachor et~al\mbox{.}}{2023}]%
        {tzachor2023large}
\bibfield{author}{\bibinfo{person}{Asaf Tzachor}, \bibinfo{person}{Medha Devare}, \bibinfo{person}{Catherine Richards}, \bibinfo{person}{Pieter Pypers}, \bibinfo{person}{Aniruddha Ghosh}, \bibinfo{person}{Jawoo Koo}, \bibinfo{person}{S Johal}, {and} \bibinfo{person}{Brian King}.} \bibinfo{year}{2023}\natexlab{}.
\newblock \showarticletitle{Large language models and agricultural extension services}.
\newblock \bibinfo{journal}{\emph{Nature food}} \bibinfo{volume}{4}, \bibinfo{number}{11} (\bibinfo{year}{2023}), \bibinfo{pages}{941--948}.
\newblock


\bibitem[\protect\citeauthoryear{Van~Zyl}{Van~Zyl}{2019}]%
        {van2019evaluation}
\bibfield{author}{\bibinfo{person}{Johannes~Gideon Van~Zyl}.} \bibinfo{year}{2019}\natexlab{}.
\newblock \emph{\bibinfo{title}{Evaluation of adjuvants in fungicide spray application for the control of alternaria brown spot in South African citrus orchards}}.
\newblock \bibinfo{thesistype}{Ph.D. Dissertation}. \bibinfo{school}{Stellenbosch: Stellenbosch University}.
\newblock


\bibitem[\protect\citeauthoryear{Xu, Li, Chen, Wang, Gao, Cai, and Ling}{Xu et~al\mbox{.}}{2024}]%
        {xu2024device}
\bibfield{author}{\bibinfo{person}{Jiajun Xu}, \bibinfo{person}{Zhiyuan Li}, \bibinfo{person}{Wei Chen}, \bibinfo{person}{Qun Wang}, \bibinfo{person}{Xin Gao}, \bibinfo{person}{Qi Cai}, {and} \bibinfo{person}{Ziyuan Ling}.} \bibinfo{year}{2024}\natexlab{}.
\newblock \showarticletitle{On-device language models: A comprehensive review}.
\newblock \bibinfo{journal}{\emph{arXiv preprint arXiv:2409.00088}} (\bibinfo{year}{2024}).
\newblock


\end{thebibliography}

\end{document}